# EUROPEAN STRATEGY FOR ACCELERATOR BASED NEUTRINO PHYSICS

*Prepared by the program committee of the Neutrino 'town meeting', CERN, 14-16 May 2012:*
Sergio Bertolucci (CERN), Alain Blondel[1] (DPNC, Geneva), Anselmo Cervera (IFIC, Valencia), Andrea Donini (IFT-UAM, Madrid), Marcos Dracos (IPNS, Strasbourg), Dominique Duchesneau (LAPP, Annecy), Fanny Dufour (DPNC, Geneva), Rob Edgecock (STFC, RAL), Ilias Efthymiopoulos (CERN), Edda Gschwendtner (CERN), Yury Kudenko (INR, Moscow), Ken Long (ICL, London), Jukka Maalampi (Jyväskylä), Mauro Mezzetto (INFN, Padova), Silvia Pascoli (IPP, Durham), Vittorio Palladino (Napoli), Ewa Rondio (Warsaw), Andre Rubbia (ETH Zurich), Carlo Rubbia (CERN), Achim Stahl (RWTH, Aachen), Luca Stanco (INFN Padova), Jenny Thomas (UCL London), David Wark (ICL & RAL), Elena Wildner (CERN), Marco Zito(CEA Saclay).

**Summary**

Massive neutrinos reveal physics beyond the Standard Model, which could have deep consequences for our understanding of the Universe. Their study should therefore receive the highest level of priority in the European Strategy. Among the many neutrino questions that experiments in different physics domains can answer, the discovery and study of leptonic CP violation and precision studies of the transitions between neutrino flavours require high intensity, high precision, long baseline accelerator neutrino experiments. The community of European neutrino physicists involved in oscillation experiments works on ongoing accelerator based experiments from CERN (CNGS), but also in Japan (T2K), the USA (MINOS), using reactors (Double Chooz) or natural sources (ANTARES, ICECUBE, km$^3$, LVD) and has taken a leading role in detector and accelerator studies towards powerful future long baseline facilities. It is strong enough to support a major neutrino long baseline project in Europe, and has an ambitious, competitive and coherent vision to propose.

---

[1] Contact Alain.Blondel@cern.ch

Since the 2006 European Strategy for Particle Physics (ESPP), the experimental situation has evolved. The $\theta_{13}$ mixing angle is now measured, with a value close to the previously available upper limits; the next steps, measurement of the mass hierarchy and of the CP violation can therefore be evaluated quantitatively. Following the 2006 ESPP recommendations, two complementary design studies have been carried out: LAGUNA/LBNO, focused on deep underground detector sites, and EUROnu, focused on high intensity neutrino facilities.

LAGUNA LBNO recommends, as first step, a conventional neutrino beam CN2PY from a CERN SPS North Area Neutrino Facility (NANF) aimed at the Pyhasälmi mine in Finland situated at 2285 km, with a Liquid Argon magnetized iron hybrid detector; this site is also adequate for deep underground physics with a high purity liquid scintillator (LENA) detector; an Expression of Interest with large community support has been submitted to this effect, and aims at a full proposal by 2014.

A sterile neutrino search experiment which could also be situated in the CERN north area has been proposed (ICARUS-NESSIE) using a two detector set-up, allowing a definitive answer to the 20 year old question open by the LSND experiment.

EUROnu has compared i) a conventional super-beam from the CERN HP-SPL, ii) a beta-beam implemented on the CERN site, and iii) a Neutrino Factory. It concluded that a 10 GeV Neutrino Factory, aimed at a magnetized neutrino detector situated, also, at a baseline of around 2200 km ($\pm$30%), would constitute the ultimate neutrino facility; it recommends that the next 5 years be devoted to the R&D, preparatory experiments and implementation study, in view of a proposal before the next ESPP update.

***The coherence and quality of this program calls for the continuation of neutrino beams at CERN after the CNGS, and for a high priority support from CERN and the member states to the experiments and R&D program.***

**Foreword**

This document results of a networking activity among the accelerator based neutrino community in Europe within the FP6/7 programs CARE (BENE) and EUCARD (NEU2012) [1] since the last European Strategy process in 2006. The 2006 strategy document stated that *'Studies of the scientific case for future neutrino facilities and the R&D into associated technologies are required to be in a position to define the optimal neutrino program based on the information available in around 2012'.* Two design studies were performed during this period.

1) EUROnu [2] has concentrated on three different neutrino facilities: a Superbeam based on the 4 MW CERN superconducting proton linac (SPL) to Fréjus (130km), a beta-beam based on the CERN accelerators and a Neutrino Factory based on a muon storage ring, studied within the framework of the International Design Study for the Neutrino Factory. EUROnu featured limited detector studies but performed a careful performance comparison between the three options.

2) LAGUNA [3] concentrated on the study of seven deep underground laboratories in Europe offering detector sites for long baseline facilities and astroparticle physics; LAGUNA was concluded in 2011 and followed by LAGUNA-LBNO (2011-2014) which narrowed the studies to three site locations and the studies of the possible conventional neutrino beams and detector set-ups for long baseline neutrino oscillation experiments.

At the same time several neutrino experiments with large European involvement have taken place, OPERA and ICARUS at CNGS, MINOS, T2K, D-Chooz, and several others; a key oscillation parameter $\theta_{13}$ was measured to be $\sin^2 2\theta_{13} \approx 0.09 \pm 0.01$, at the top of the previously allowed range, changing the overall picture significantly. The European neutrino community was convened at the occasion of several workshops [4][5][6][7], aiming at defining the future strategy, both by making the best use of existing facilities and designing the ultimate facility. Two of these gatherings were organized in collaboration with the CERN management, with participation of up to 250 participants. The emphasis was placed on understanding the possible synergy of the LAGUNA and EUROnu studies, which could be found along two different paths: 1) the short baseline path based on the CERN to Fréjus baseline, with a SPL based superbeam aimed at a 500 kton Water Cherenkov with a $\gamma$=100 beta-beam as long term vision[2]; 2) the long baseline path, based on the CERN to Pyhasälmi baseline with a SPS-based superbeam as a first step, and ultimately optimal for a Neutrino Factory. While the first path offers a powerful reach on CP violation and underground physics, it has a high entry price since the SPL, although it remains one of the R&D lines at CERN, is no longer part of the LHC upgrade plan; the second approach has a smaller entry price and more precise ultimate possibilities; **the long baseline path is preferred by both design study collaborations, thus offering a coherent long term vision.**

The possibility to revive a sterile neutrino search was also considered. From the original proposal referring to a refurbished PS neutrino beam facility, the study has evolved towards a beam from the SPS, possibly sharing the extraction elements, allowing a better synergy between short, middle and long term plans.

The concluding 'town' meeting, 'European Strategy for future neutrino physics II' was held at CERN 14-16 May 2012 [7]; what is presented below is based on the material presented, and the discussion organized at the occasion of this workshop. It was drafted by the program committee in which, as can be seen, took part proponents of the main projects under discussion.

---

[2] The SPL and accumulator of the EUROnu superbeam could also serve as basis for a Neutrino Factory proton driver.

# 1. The importance of Neutrino Physics

The existence of massive neutrinos is, today, the only experimental demonstration of new physics beyond the Standard Model of particle physics (SM) [8]. It requires extension of the SM in a way that is yet unknown. Neutrino masses and mixing angles may offer insight into the poorly understood question of the generation of particle masses, and are expected to reveal leptonic CP violation. The existence of a Dirac mass term would imply the existence of new right-handed neutrinos, while the existence of a Majorana mass term would imply fermion number violation. Neither mass term being forbidden by experimentally verified conservation laws, the theoretically favored scenario involves both, leading to the existence of right-handed (a.k.a. 'sterile') neutrinos with masses different from the left-handed (a.k.a. 'active') ones within a very large range of possible masses extending from 0 to almost the GUT scale. This situation opens a deep and promising field of research, with a large potential for discoveries of considerable consequences. The following (in no particular order of importance) would constitute high level discoveries:

1. determination of the absolute mass scale of neutrinos
2. observation of an inverted mass hierarchy of the active neutrinos
3. CP violation in neutrino oscillations
4. violation of unitarity of the neutrino mixing matrix
5. observation of neutrinoless double beta decay demonstrating fermion number violation
6. discovery of effects implying unambiguously the existence of sterile neutrino(s).

The potential consequences of such discoveries would be to offer insight on the nature of particle masses and the question of flavor; CP violation in conjunction with fermion number violation could lead to a rather natural explanation of the origin of the matter-antimatter asymmetry of the Universe by Leptogenesis. Sterile neutrinos offer very natural dark matter candidates; the neutrino mass scale, the knowledge of the neutrino spectrum and the possible existence of light sterile neutrinos ($eV/c^2$ scale) are essential ingredients in the understanding of the evolution of the early Universe. In addition, the deep underground detectors offer powerful physics synergies with e.g. proton decay searches, detection of geophysical, solar and atmospheric neutrinos and neutrinos from other astrophysical sources such as supernovae.

These measurements require a variety of different techniques, which deserve to be given high priority across the board. ***Discovery and study of CP violation can only be done with accelerator based neutrino beams,*** *which* can also contribute in a decisive way to the determination of the mass hierarchy, precision studies of the neutrino mixing matrix and its unitarity and the search for sterile neutrinos.

***Europe should place neutrino physics at the highest level of priority.***

## 2. The neutrino landscape and the next steps

The recent measurement of $\sin^2 2\theta_{13} \sim 0.09 \pm 0.01$ by reactor experiments [9] and the observation of $\nu_\mu \to \nu_e$ appearance in long baseline experiments [10] clarify the next steps to follow. The relatively large value of $\theta_{13}$ and of the appearance rate will allow a clear-cut determination of the mass hierarchy of neutrinos, and may render some observation of CP violation accessible, though not easily, with conventional neutrino beams, within the next 10-20 years depending on the value of parameters.

New and better defined neutrino beams from the Neutrino Factory will become necessary for precise and definitive study of leptonic CP violation, a full verification of the 3x3 mixing of active neutrinos and the search for physics beyond this framework such as mixing with sterile neutrinos or some unexpected surprise, by precise determination of all possible flavour transitions of neutrinos.

The search for sterile neutrinos is an extremely broad field as their masses are not constrained between few meV to $10^9$ GeV, and can be pursued in a great variety of means. A good starting point is the clarification of the possible anomalies in nuclear reactor and short baseline experiments, using nuclear sources and short baseline accelerator experiments

***The determination of the neutrino mass hierarchy and the determination of the CP phase are the next steps in long baseline neutrino experiments. These fundamental measurements require and justify dedicated long baseline accelerator-based experiments.***

Triggered by the relatively large value of $\sin^2 2\theta_{13}$, a number of possibilities have been brought up to determine of the mass hierarchy in the medium time scale. These are summarized in the Table **1**. The sensitivity to the mass hierarchy comes from the observation of the modulation of the solar oscillation by the atmospheric oscillation in case of the 60 km reactor experiment, while for most of the other projects it originates from the difference in electron vs antineutrino electron appearance due to matter effects in the earth for experiments using a very long baseline. While it is possible that a combination of the already approved experiments (for instance T2K +NOvA + the reactor experiments) will be able to achieve a separation of 3σ between the inverted and 'normal' hierarchy, this is not guaranteed as it will depend on the value of the CP violating phase δ. The projects that are certain to provide a definitive determination (5σ) are LBNO, the Neutrino Factory and, possibly, using atmospheric neutrinos in an extension of ICECUBE called PINGU, which however remains to be demonstrated. From all the proposed ideas, LBNO thanks to its very long baseline, will be able to verify in detail the phenomenon of matter effects, testing the prediction of three neutrino flavor mixing, and thereby determine the mass hierarchy. This is different from the extraction of the hierarchy from the combined fit of several experimental observations, which assumes the correctness of the model.

Table 1 Claimed sensitivities on the determination of the mass hierarchy. There are considerable uncertainties on the performance, feasibility and time scale of many of these projects. Approved at the moment are MINOS+, NOVA, ICAL@INO; LOI or EOI have been produced for HyperK [11] and LBNO [12]. Most of the numbers in the table are gathered from presentations at the Neutrino 2012 conference or at the Neutrino Town meeting at CERN; they are the responsibility of their authors.

| Project | | | Separation of IH and NH | Pre-requisite and date of achievement | Reference |
|---|---|---|---|---|---|
| DayaBay II | reactor 60km | 20 kt LS | 3 σ in 6 years | R&D on E-resolution 2020 ? | Karsten Heeger at Neutrino 2012 |
| ICAL@INO | atmospherics | 50 kt MID (RPCs) | 2.7 σ in 10 years | 2027 | Sandhya Choubey at Neutrino 2012 |
| HyperK | atmosherics | 1 Mt Water Cerenkov | 3 σ in 5 years 4 σ in 10 years | 2027/28 2033/34 | HyperK LOI Sandhya Choubey at Neutrino 2012 |
| T2HK | LBL accel. 295 km | 1 Mt Water Cerenkov | 0..3 σ in 10 years | 2028 | Masashi Yokoyama at Neutrino 2012 |
| PINGU | atmospheircs | Ice (South pole) | 3...11 σ in 5 years | feasibility study ongoing, understanding of resolution and systematics on atmospherics Around 2020 if it works. | Uli Katz at neutrino Town meeting |
| MINOS+ | LBL accel. 735 km | MID 5.4 kt | no claim on mass hierarchy | --- | Jenny Thomas at neutrino Town meeting |
| GLADE | LBL accel. 810 km | LAr 5 kt | In combination with NOvA and T2K: ≤ 2 σ | Letter-of-Intent | |
| NOvA | LBL AshRiver 810 km | TASD 14 kt | 0...3 σ in 6 years depending on δ | Full operation in 2014 2020 | Ryan Patterson at Neutrino 2012 |
| LBNE | LBL Homestake LBL Soudan LBL AshRiver | LAr 10 kt LAr 15 kt LAr 30 kt | 1.5...7 σ in 10 y 0...3 σ in 10 y 0.5...5 σ in 10 y | 2030 | Bob Swoboda at Neutrino 2012 |
| LBNO | LBL accel. 2300 km | LAr 20 kt | > 5σ in a few y. | 2023 + If decision in 2015 | André Rubbia at Neutrino 2012 |
| LENA | LBL accel. 2300 km | Liq. Scint. 50 kt | 5 σ in 10 years | 2028 + number of years to the decision | Lothar Oberauer at Neutrino 2012 |
| Neutrino Factory | LBL accel. 2000 km | MIND 100kton | >> 5σ | | Ken Long at Neutrino 2012 |

The observation of the CP violation and the determination of the CP violating parameter $\delta$ is performed by analysis of the appearance reaction $\nu_\mu \rightarrow \nu_e$ (once the value of $\theta_{13}$ is known and the mass hierarchy determined, $\delta$ is the sole remaining parameter) or of the asymmetry:
$A_{CP} = (P(\nu_\mu \rightarrow \nu_e) - P(\bar{\nu}_\mu \rightarrow \bar{\nu}_e))/(P(\nu_\mu \rightarrow \nu_e) + P(\bar{\nu}_\mu \rightarrow \bar{\nu}_e))$.

Now that the angle $\theta_{13}$ is known, the performance of various experiments on the measurement of the CP phase delta can be compared univocally. This was done in the framework of EUROnu [13]. The projects have been compared under the following assumptions: the T2HK project (2.5$^0$ off axis beam, 295 km, 500 kton of Water, 4MW of beam power for 8 years); the LBNE project (wide band beam, 1290 km, 33kton of Liquid Argon TPC, 700kW of beam power for 10 years); the SPL project (Wide band beam, 130 km, 440kton Water Cherenkov, 4MW beam power for 10 years); C2P (2300 km, 100kton Liquid Argon TPC, 800kW of beam power for 10 years); the γ=350 beta-beam (requiring a new 1TeV SPS at CERN) , it being observed that a realistic γ=100 beta-beam has worse performance than the SPL; and finally the 10 GeV Neutrino Factory aiming at a 100 kton magnetized iron detector on a 2000 km baseline. Clearly the Neutrino Factory provides the most sensitive determination for all values of δ. The experimental accuracy of T2HK, keeping in mind the optimistic assumptions of the study, is quite interesting, it benefits from the fact that this experiment has internally a 3 sigma sensitivity on the mass hierarchy with atmospheric neutrinos. Given the unavoidable uncertainties and variations in the assumptions made for these estimates, the choice between the super-beam projects becomes more a question of opportunity and planning. Together with the evaluation of the relative cost of these facilities, the EUROnu Design Study concluded that the Neutrino Factory should be developed as the most powerful facility for precision study of neutrino oscillations.

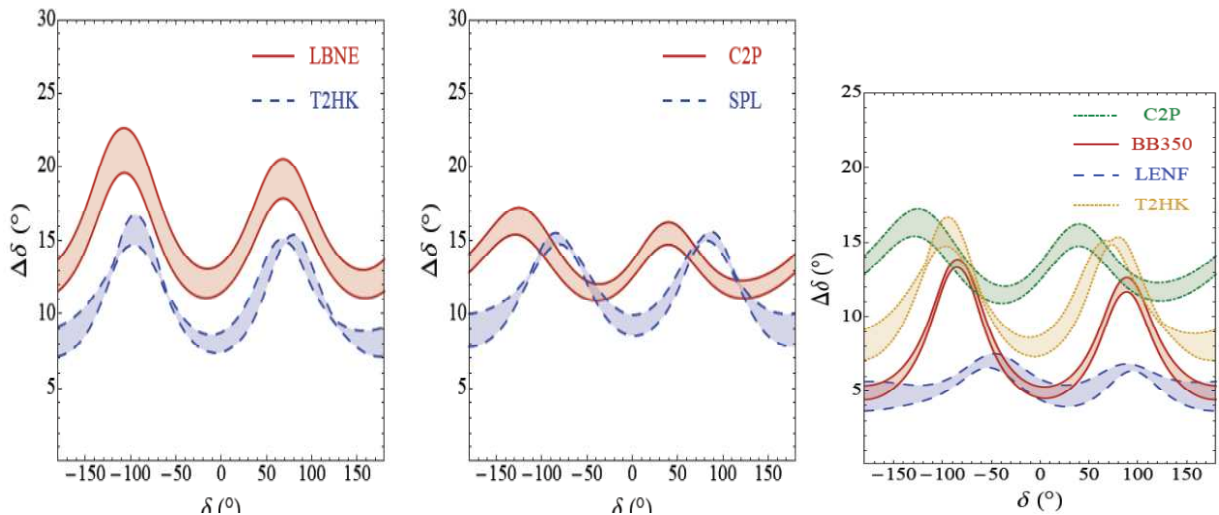

Figure 1 comparison of CP sensitivity and precision on the CP violating phase δ of different facilities [13].

## 3 Community and global context

The neutrino oscillation community involved in running experiments addressing neutrino oscillations in Europe is not small. A census organized at the occasion of the town meeting counted important groups working on the CNGS experiments OPERA (180) and ICARUS(60),  on DChooz(170), T2K(250), MINOS (25),  BOREXINO (100), ANTARES(153) and  ICECUBE (125) where the numbers in parenthesis indicate the European particle physicists. A total number of around

750 can be estimated at present. The European physicists are playing a leading role in R&D and studies towards future accelerators and detectors (MERIT, MICE, Beta beam, IDS-NF, EUROnu, AIDA and LAGUNA) and in the ancillary experiments necessary for the precision of accelerator-based experiments worldwide (NA61/SHINE). At CERN itself there is a strong accelerator group on the CNGS – but no physics group at the moment. It has been argued that building a neutrino physics team at CERN with a reasonable size project achievable in the next 10 years is a prerequisite to more ambitious long term goals.

The presently running or approved experiments in the US (MINOS+, NOVA) and in Japan (T2K) will improve the measurements of oscillation parameters considerably, but are not certain to be able to carry out the determination of MH and discovery of CPV in a definitive way (5$\sigma$). Further experiments (LBNE, T2HK) are being discussed but their exact scope and time scale is not clear yet. The projects are largely complementary, especially T2HK (300km) with the longer baseline options, and the importance of the subject justifies a competitive approach.

## 4. Opportunity for the next step in Europe.

The next step should be an experiment which can be constructed in a reasonable time (less than about 10 years), maintains the community healthy, with a real chance of discovery and long term upgrade possibilities. The existence of a possible long baseline in Europe from CERN to Pyhasalmi (2300 km) is unique in this regard. The Pyhasalmi mine was selected by the LAGUNA consortium following the initial design study (see [12] for more details).

Building on the experience with CNGS and on the pioneering competence in Liquid Argon TPCs, European physicists are in the position to propose a realistic next step: a conventional neutrino beam CN2PY in the CERN North Area Neutrino Facility (NANF), under study, aiming at 20kton of fine grain detector (GLACIER) followed by a 40 kton magnetized iron detector (MIND) at Pyhasalmi. It should be supported by extensive hadroproduction measurements in an upgraded SHINE/NA61 and a near detector. This project, called LBNO, is the first priority of the LAGUNA-LBNO consortium and is endorsed by the Neutrino Factory community. An expression of Interest [12] has been submitted to the CERN SPSC with ~240 signatures from European institutions.

LBNO can study the oscillatory behavior as a function of a wide range of energies independently for neutrinos and antineutrinos, and will achieve a definitive determination of the neutrino mass hierarchy quite rapidly: 2 years at present CNGS intensity for a $\geq$3$\sigma$ determination for any value of $\delta$; if the SPS delivered power can be raised to 750kW, the experiment will reach 5$\sigma$ determination in that time. The experiment will offer a first level of sensitivity to CP violation and studies of oscillations into tau neutrino (at that distance, the first oscillation maximum is at 4.6 GeV and thus above the tau threshold) and search for sterile neutrinos by neutral current disappearance.

The deep underground location allows non accelerator applications, such as described in the LBNO EoI or those provided in addition by the 50kton liquid scintillator LENA project. Both the quality of the underground facilities and the distance to CERN make it such that it can evolve into a larger detector and a more powerful beam from a Neutrino Factory and thus, offers a long term vision.

**Discussion** The quality and possible shortcomings of the LBNO project were discussed extensively and critically at the neutrino 'town' meeting. Concerns were expressed in particular on the following aspects. (More detailed answers can be found in the LBNO EOI).

*-- there is concern that because of the ~10% slope of the beam, a near detector station may not be feasible with the desired performance.* It is planned to have a magnetized near detector based on a Argon gas TPC + magnetized iron at a distance between 500 m and 800 m from the target. The depth is similar to that of e.g. the LEP or LHC experiments, and should not prevent performing high quality measurements.

*-- the long baseline introduces matter effect that allows a very sensitive determination of the mass hierarchy, but the associated uncertainty affects the CP violation measurement.* This uncertainty has been estimated (see the EOI for a complete discussion) and was taken into account in all LBNO and Neutrino Factory studies. It is not a dominant systematic error until one reaches an uncertainty of a few degrees on the phase $\delta$.

*-- the tau production followed by semileptonic decays results in a background to the $\nu_\mu \rightarrow \nu_e$ 'appearance channel' and may affect the sensitivity – especially of the second maximum of oscillation at Ev=1.5 GeV.* The tau production is actually a very interesting physics effect in its own right; detection of the non-leptonic decays of the taus will allow a direct calculation of the leptonic ones and thus of the background.

*-- The second oscillation maximum requires extremely good control of energy reconstruction that has not been demonstrated so far.* This is completely true and stresses the importance of a dedicated test beam effort for study of precise reconstruction of hadronic interactions in a Larg TPC or a magnetized iron calorimeter.

*-- Mass hierarchy effects are so big that either the neutrino or the antineutrino beam (depending from the hierarchy) will be almost fully suppressed. CP violation would be extracted from fits to sin($\delta$) and not by $\nu - \bar{\nu}$ asymmetry. Note that even $\nu_\mu$ disappearance can provide fits to cos($\delta$).* This point was studied in the LBNO EOI, but applies also to the Neutrino Factory on the same baseline. The available channels indeed provide sensitivity to CP violation, and a measurement of both sin$\delta$ and cos$\delta$. Similarly, to T2HK or CERN-Fréjus, the running time in neutrino and antineutrino modes will be optimized once the mass hierarchy is known (see EoI).

It was agreed that further studies will be requested before a complete proposal can be made, in 2014. It is generally felt that, because of the long distance, the oscillation physics at LBNO involving matter effects and tau production is both more complex and richer than that of a simple $v_\mu \rightarrow v_e$ appearance experiment.

## 5 Short baseline neutrino beam and sterile neutrino search

The search for sterile neutrinos is an extremely broad field as their masses are not constrained between few meV to $10^9$ GeV, and can be pursued in a great variety of means, involving an apparent violation of the results predicted with only 3x3 mixing of active flavours [14].

A good starting point is the clarification of the possible anomalies observed [14] in nuclear reactor, radioactive sources, short baseline experiments and the number of effective neutrinos flavors in cosmology (Neff~4). None of these observations is truly convincing methodologically, they conflict with a number of other experimental results and the whole situation is not fit very well to a coherent model with sterile neutrino(s). However the anomalies are felt to be strong enough to demand clarification with a methodologically sound experiment [16].

An experimental search for sterile neutrinos at a new CERN-SPS neutrino beam has been recently proposed by the ICARUS-NESSiE Collaboration (SPSC-P347, 150 physicists) [15]. The experiment is based on two identical LAr-TPCs complemented by magnetized spectrometers, observing the electron and muon neutrino events at the far and near positions 1600 and 300 m from the proton target, respectively. The project will exploit the ICARUS T600 detector, the largest LAr-TPC ever built with a size of about 600 t of imaging mass moved to the CERN Far position with an additional clone of 150 t of LAr located in the near position. Two magnetic spectrometers will be placed down-stream of the two LAr-TPC detectors to greatly complement the physics capabilities. The proposed experiment will explore in a definitive way a region of parameter space completely covering the possible anomaly claimed by LSND and a large fraction of the region relevant to the reactor anomalies. Looking for the electron neutrino oscillation in both appearance and disappearance modes, as well as the muon neutrino Charged Current and Neutral Current disappearance will allow disentangling the different possible couplings to sterile neutrinos and cover all possible light sterile neutrino signatures for masses up to tens of ev$^2$. In addition, given the similar concept of these detectors with those considered for the Long Baseline Neutrino Oscillation CN2PY experiment, there is an evident synergy in the technologies.

A number of other projects reviewed in [14] are under discussion to address this question but none addresses the 'LSND anomaly' as directly and completely as the ICARUS-NESSIE proposal. To concentrate on projects involving high energy accelerator neutrinos, the MINOS+ experiment running with the Fermilab NUMI beam at medium energy will explore an overlapping part of the parameter space [17]. A project similar to [15] is discussed using microBooNE and a 1kton LAr detector on the miniBooNE beamline [18], however both detector event rates are significantly lower than in the CERN proposal, and the experiment does not include a magnetic spectrometer to study muon neutrino disappearance.

Another goal of the ICARUS-NESSIE proposal is the precise measurement of cross-sections, requiring a well known neutrino flux. Since the requested beam would be generated using protons accelerated in the SPS to 100 GeV, the necessary hadroproduction measurements could readily be made within the NA61/SHINE experiment with a precision of 3-5%.

Although it is also in the north area, the beamline for the ICARUS-NESSIE proposal is not identical to that of the LBNO project. ***Consistency in time and protons with the Long Baseline Neutrino Oscillation program outlined in the previous section must be ensured.***

Finally it is interesting to note the nuSTORM project (at the stage of an LOI) at Fermilab [19]. It is based on neutrinos from a muon storage ring, a precursor to a Neutrino Factory. Certainly nuSTORM could perform very powerful short baseline oscillation studies; furthermore, the outstanding knowledge of the neutrino flux could lead to cross-section measurements for all four available flavours of neutrinos at a new level of accuracy (percent or better), necessary for the precise determination of CP violation in the superbeam experiments.

## 6. Preparing for longer term, precision experiments

Given the large value of θ13, the conventional long-baseline experiments will determine the neutrino mass hierarchy. However, their sensitivity to leptonic CP-invariance violation over the whole range of δCP might not be sufficient to uncover leptonic CP violation with high significance and redundancy. Furthermore, to establish the consistency of the 3x3 neutrino mixing Standard Neutrino Model thoroughly and to test theories that purport to explain the physics of flavour will require measurements with a precision beyond that which the next generation of experiments can deliver. It is therefore essential to prepare the ground for a proposal for a facility capable of making measurements of the requisite precision.

The EUROnu Design Study has studied and compared three facilities proposed to serve the era of precision measurements: a super beam, the Neutrino Factory and the beta-beam. After consideration of the relative cost, complexity, timescale and interference with the exploitation of the LHC, EUROnu concluded that a Neutrino Factory based on a 10 GeV muon storage ring aimed at a distance of 2200 km (with a broad optimum of ±30%) is the facility of choice to serve the precision era. This is the same baseline as that from CERN to Pyhasalmi in Finland proposed by the LBNO collaboration, thus taking advantage of the existence and preparation of the far detector laboratory and offering a strong synergy between the two projects.

**The European Strategy for Particle Physics must therefore provide for European participation in the programme required for a Neutrino Factory proposal to be prepared in time for the next update of the European Strategy.**

This programme must encompass:

-- The completion of the necessary hardware and system R&D including the MICE experiment;

-- The experimental demonstration that stored muons can serve a first-rate neutrino programme with the percent precision measurement of the $\nu_{e,\mu}$ N and $\bar{\nu}_{e,\mu}$ N scattering cross sections and the search for sterile neutrinos using the nuSTORM experiment ;

-- All relevant design work, including consideration of the implementation of the facility at CERN.